\documentclass{aa}  

\usepackage{graphicx,hyperref}
\newcommand{\ktfour}{KELT-24}
\newcommand{\fxu}{erg s$^{-1}$ cm$^{-2}$}
\newcommand{\lxu}{erg s$^{-1}$}
\newcommand{\xmm}{{\sl XMM-Newton}}
\newcommand{\pn}{{\sl pn}}
\newcommand{\logrhk}{ $\log R^\prime_{HK}$ }
\usepackage{txfonts}
\begin{document}

\title{The X-ray activity of F stars with hot Jupiters: KELT-24 versus WASP-18}


\author{I. Pillitteri\inst{1}
          \and
        S. Colombo\inst{1}
        \and
        G. Micela\inst{1}
          \and
        S. J. Wolk\inst{2}    
          }

   \institute{INAF-Osservatorio Astronomico di Palermo, Piazza del Parlamento 1, 90134 Palermo, Italy\\
              \email{ignazio.pillitteri@inaf.it}
         \and
           Harvard-Smithsonian Center for Astrophysics, 60 Garden St, Cambridge (MA) 02138, USA 
             }

   \date{Received; accepted }

 
  \abstract
   {X-rays emitted by the coronae of solar-type stars are a feature present in up to late-A types during the main sequence phase. 
   F stars, either with or without hot Jupiters, are usually X-ray emitters. 
    The very low level of X-ray emission of the F5 star WASP-18 despite its 
    relatively young age and spectral type is thus quite peculiar. 
    In this paper we compare the X-ray activity of \ktfour\ to that of WASP-18. \ktfour\ is an F5 star 
    nearly coeval to the Hyades stars that hosts a 5 $M_\mathrm{Jup}$ in a 5.6-day period orbit. 
    The properties of the \ktfour\ system are similar to, although less extreme than,
    those of WASP-18. 
    We observed \ktfour\ with \xmm\ for a total of 43 ks in order to
    test if the X-ray activity of this star is depressed by the interaction with its massive hot 
    Jupiter, as is the case of WASP-18. 
    \ktfour\ is detected in combined EPIC images with a high significance level.
    Its average coronal spectrum is well described by a cool component at 0.36 keV and a 
    hotter component at 0.98 keV. 
    We detected a flare with a duration of about 2 ks, during which the coronal temperature reached 
    3.5 keV. The unabsorbed quiescent flux in 0.3-8.0 keV is $\sim1.33\times10^{-13}$ \fxu, 
    corresponding to a luminosity of $1.5\times10^{29}$ \lxu at the distance of the star. 
    The luminosity is well within the range of the typical X-ray luminosity of F stars in Hyades,
    which are coeval. We conclude that the activity of \ktfour\ appears normal, as expected, and is not affected by any star--planet interaction. 
    From the analysis of TESS light curves, we infer a distribution of optical flares for \ktfour\ and WASP-18.
    Small optical flickering similar to flares is recognized in WASP-18 but at lower levels of energy and amplitude than in \ktfour.  
    We discuss the causes of the low activity of WASP-18. Either WASP-18b could hamper the formation of a corona bright in X-rays in its host star 
    through some form of tidal interaction, or the star has entered a minimum of activity similar to the solar Maunder minimum. 
    This latter hypothesis would make WASP-18 among the few candidates showing such a quench of stellar activity.}


   \maketitle
%

\section{Introduction}
More than 300 exoplanets out of the about 5000 known today orbit around their stars in less than 10 days 
and have masses $>0.5$ M$_\mathrm J$, the so-called hot Jupiters (hJs). 
It is plausible that  systems with hJs manifest phenomena of star--planet 
interaction (SPI) of magnetic and/or tidal origin \citep{Cuntz2000,Shkolnik03,Shkolnik2008,Ilic2022}.
{ In} low-mass stars, SPI can produce an enhancement of the coronal emission due to the interaction 
between stellar and planetary magnetospheres that could lead to flares
\citep[see][]{Lanza2013, Klein2022}. 
In another scenario, the evaporation of the planetary atmosphere 
can form a stream of gas that could accrete onto the star \citep{Matsakos2015,Pillitteri2015,Colombo2022AN}.
Planetary material stripped from the gravitational pull of the star at the periastron passage 
could precipitate onto the stellar surface, generating soft X-rays \citep{Maggio2015}

Tidal SPI can transfer angular momentum from the planet to the star and lead to an 
enhancement in the overall stellar activity since it is linked to the stellar rotational rate. 
This is the case of well-studied systems with hJs such as HD~189733 
\citep{Pillitteri2010,Pillitteri2011,Pillitteri2014a}, HD~17156 \citep{Maggio2015}, 
Corot-2A \citep{Poppenhaeger2014},  and HD~179949 \citep{Shkolnik2008}.

Among the stars with massive hJs, the F6 type star WASP-18 is quite peculiar. 
Its planet has a mass of $\sim10$M$_\mathrm J$ and orbits in less 
than one day around the star. The star has a very low X-ray emission 
($\le 5\times10^{26}$ erg/s) as well as an almost null chromospheric activity \citep{Miller2012,Pillitteri2014b,Fossati2018}. 
The expected level of X-ray luminosity for a young F star 
such as WASP-18 is on the order of $1-2\times10^{29}$ erg/s. 
\citet{Pillitteri2014b} speculated that the strong discrepancy between the expected and measured
X-ray emission could arise from the tidal interaction of the massive hJ 
with the star and its thin convective layer.
This would prevent the establishment of a large magnetic dynamo and thus a corona. 
As a result, the star would be dark in X-rays. 
Adequate modeling supporting this scenario is still missing
and requires a detailed description that connects the inner stellar structure, 
the creation and emergence of the magnetic field, and its interaction with a 
tidal perturbation due to the planet.

A system with characteristics similar to WASP-18, although not as extreme, 
is  \ktfour\ (other designations: HD\,93148 and MASCARA-3). 
\ktfour\ is composed of a young ($\sim$780 Myr) F5-F7 star orbited by a 5.1~M$_J$ hJ every 5.55 days 
\citep{Rodriguez2019,Hjorth2019}. \ktfour\  has an effective temperature of $\sim6500$ K,
a radius of 1.52 R$\odot$, a bolometric luminosity of $L_\mathrm{bol}\sim1.43\times10^{34}$ 
\lxu, and is located at about 96.8 pc from the Sun.
The star was undetected in the ROSAT All Sky Survey (RASS; \citealp{Boller2016}) 
below a limit flux $f_{X,lim}\sim10^{-12}$ \fxu. Nevertheless, this flux is above the value expected for such a star
($f_X\sim 10^{-13}$ \fxu\ for a luminosity $L_X\sim10^{29}$ erg/s).
In this paper we report the use of \xmm\ to detect and measure the X-ray activity of 
\ktfour\ and compare its X-ray emission to that of WASP-18. 
The structure of the paper is as follows: Sect. \ref{obs} describes the observations and
the data analysis, Sect. \ref{results} shows the results of the analysis of X-ray and { in optical
band from NASA Transiting Exoplanet Survey Satellite (TESS)} 
data, and Sect. \ref{conclusions} contains our discussion and conclusions. 

\begin{figure}
\resizebox{\columnwidth}{!}{
   \includegraphics{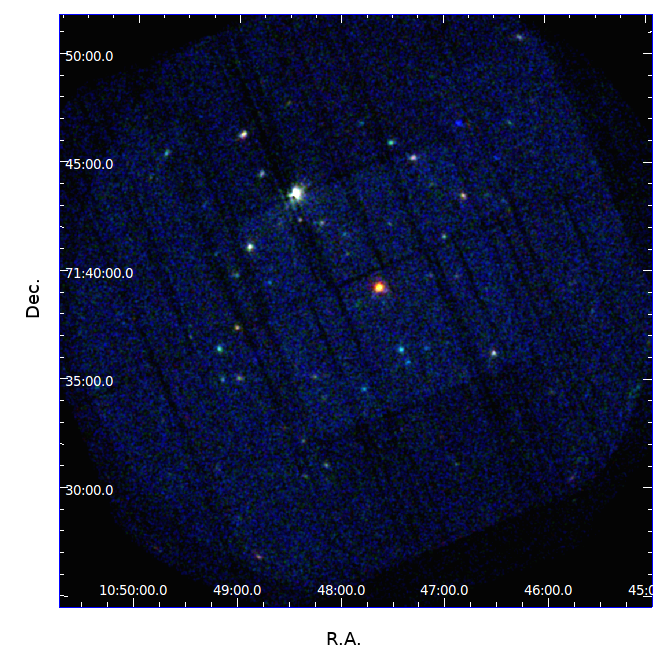}
}
\caption{\label{rgb} Combined image of \xmm\ MOS and \pn\ exposures toward \ktfour, 
which is the source at the center. The color channels correspond to: $0.3-1.0$ keV (red), 
$1.0-3.0$ keV (green), and $3.0-8.0$ keV (blue).}
\end{figure}

\section{Observations and data analysis \label{obs}}
\begin{figure}
\resizebox{\columnwidth}{!}{
   \includegraphics{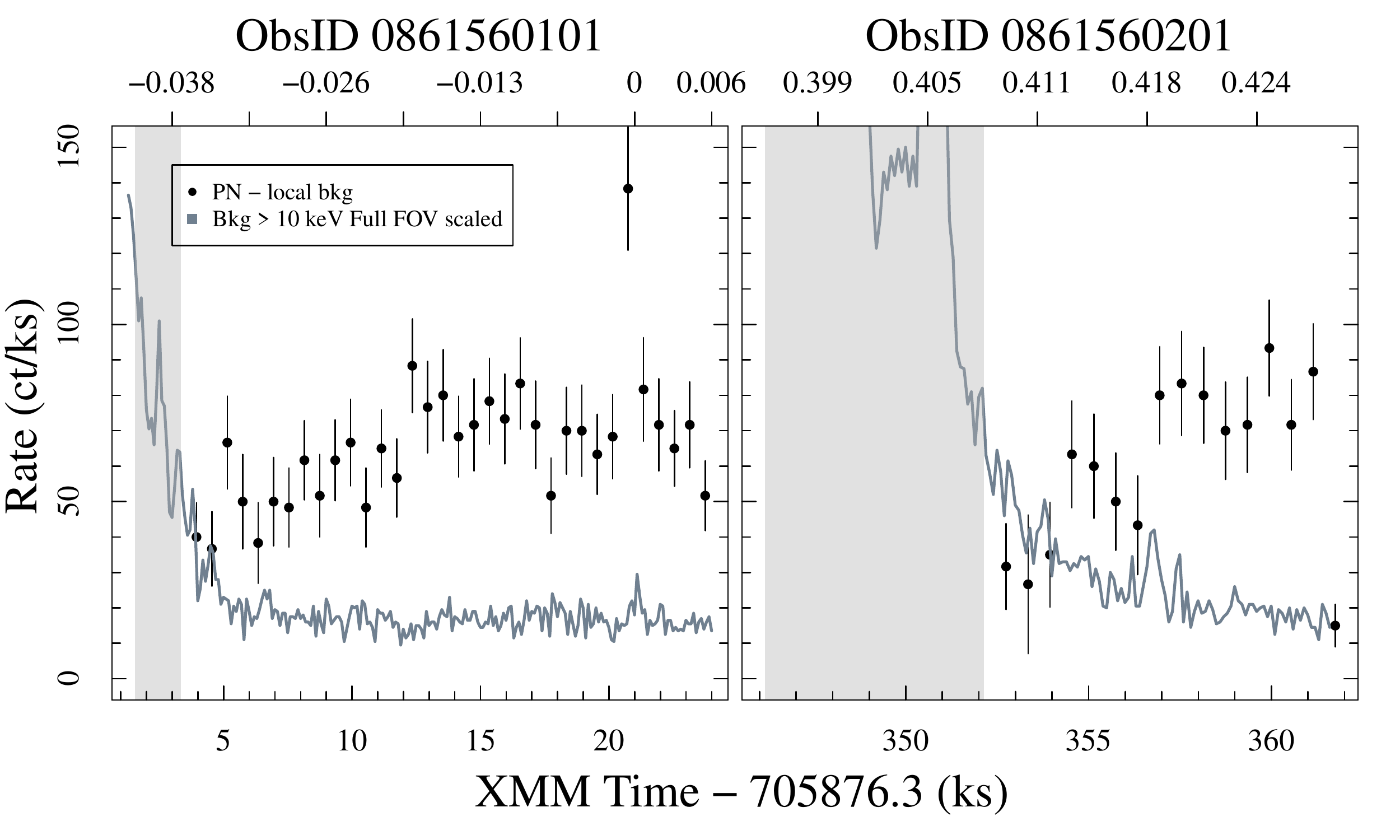}
}
\caption{\label{lc} 
 \pn\ light curves of \ktfour\ obtained in the first (left panel) and second observation (right panel).
The  two time series are the net rate of the star (black symbols) and the background at energies above 10 keV
(gray curve). Error bars at the 1 sigma confidence level are shown. 
The X-axis at the bottom indicates the time elapsed since the beginning of the first observation. 
On the top axis we mark the planetary phases of the primary transit. 
The gray areas show the time intervals of high background excluded from the analysis. }
\end{figure}

\ktfour\ was observed with \xmm\ on May 14 and May 18, 2020, for a duration of about 25 ks and 18 ks, respectively.
The main instrument was EPIC, 
used in full frame imaging mode with the medium filter. 
The nominal aim point  of observations was at the position of  \ktfour\ 
($\alpha = 10^h 47^m 38.3^s $, $\delta = +71^d 39^m 21.1^s$, J2000). 
The data sets were retrieved from the \xmm\ science 
archive\footnote{\url{https://www.cosmos.esa.int/web/xmm-newton/xsa}}, 
reduced with SAS 
version 18.0.0 and 20.0.0, and filtered to obtain tables of events in the energy band 
0.3--8.0 keV, with {\sc FLAG = 0} and {\sc PATTERN <= 12} for the MOS
and \pn\ 
cameras, as  prescribed by the SAS guide.
The two \xmm\ exposures were affected by high background during the first $\sim2$ ks and $\sim6$ ks, 
respectively (Fig. \ref{lc}).
These intervals were discarded from the subsequent timing and spectral analysis. 
We used the light curves of the events in the full field of view in the MOS and \pn\ { chips} 
with energies  above 10 keV (10--12 keV for \pn) to evaluate their variability and determine a threshold count rate.
This threshold was used to discard intervals with count rates higher
than 0.6 ct s$^{-1}$ or 1 ct s$^{-1}$ for \pn\ and 0.25 ct s$^{-1}$ or 0.5 ct s$^{-1}$ 
for MOS for the first and second exposures, respectively.

\subsection{Spectra and light curves}
Spectra and light curves of MOS and \pn\ of \ktfour\ were obtained with the SAS {\sc evselect} task 
from events accumulated in circular regions of radius 40\arcsec. 
For the local background we accumulated the events in circular regions of 60\arcsec for  MOS and 
40\arcsec for \pn\ near the position of \ktfour\ and free from other sources. For the \pn\ a further 
constraint was to select the background from a region at approximately the same distance  
from the readout node as the source region,  as prescribed by the SAS guide, 
resulting in a region with a radius of 40\arcsec for the local background. 

\begin{figure*}
\resizebox{\textwidth}{!}{
   \includegraphics{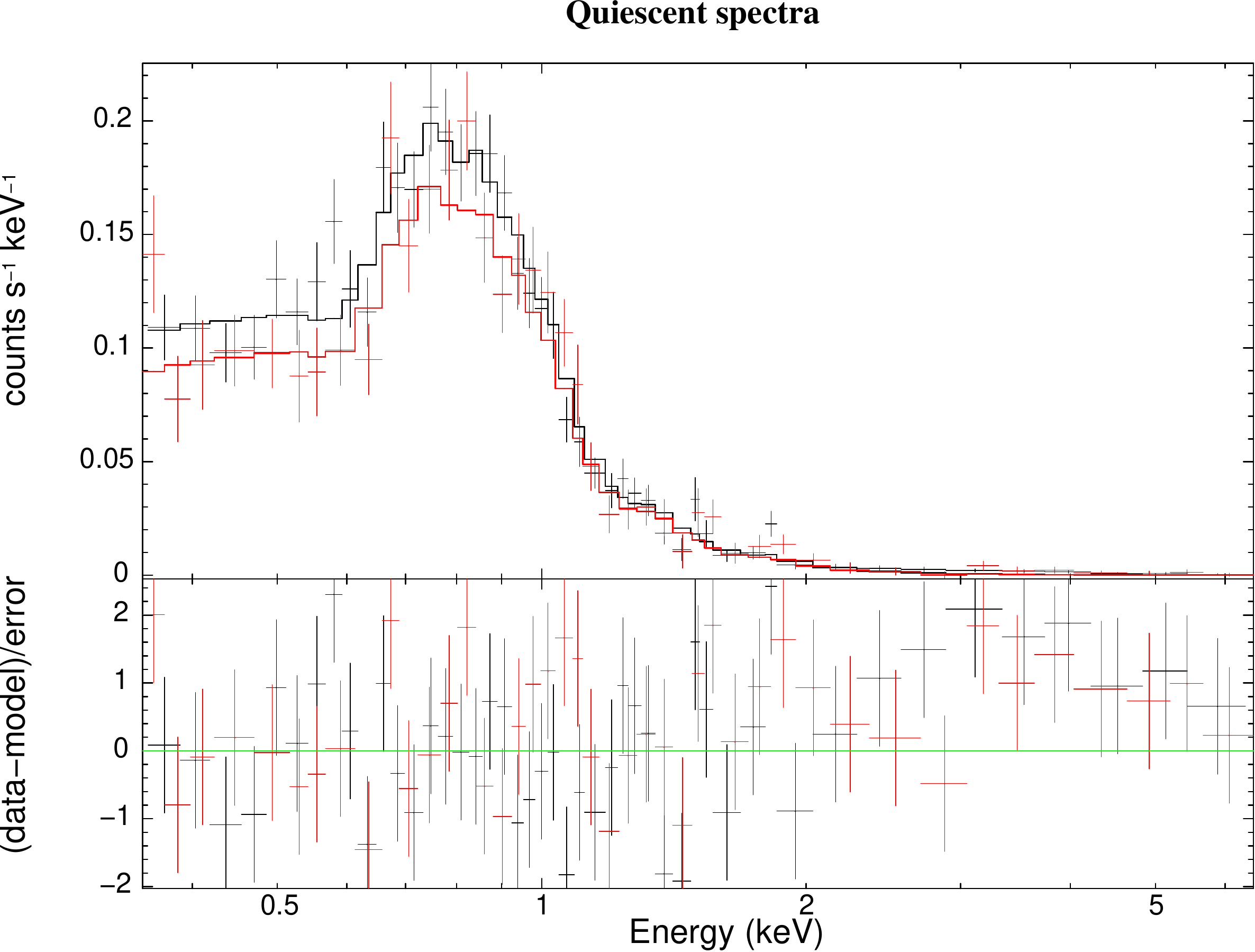}
   \includegraphics{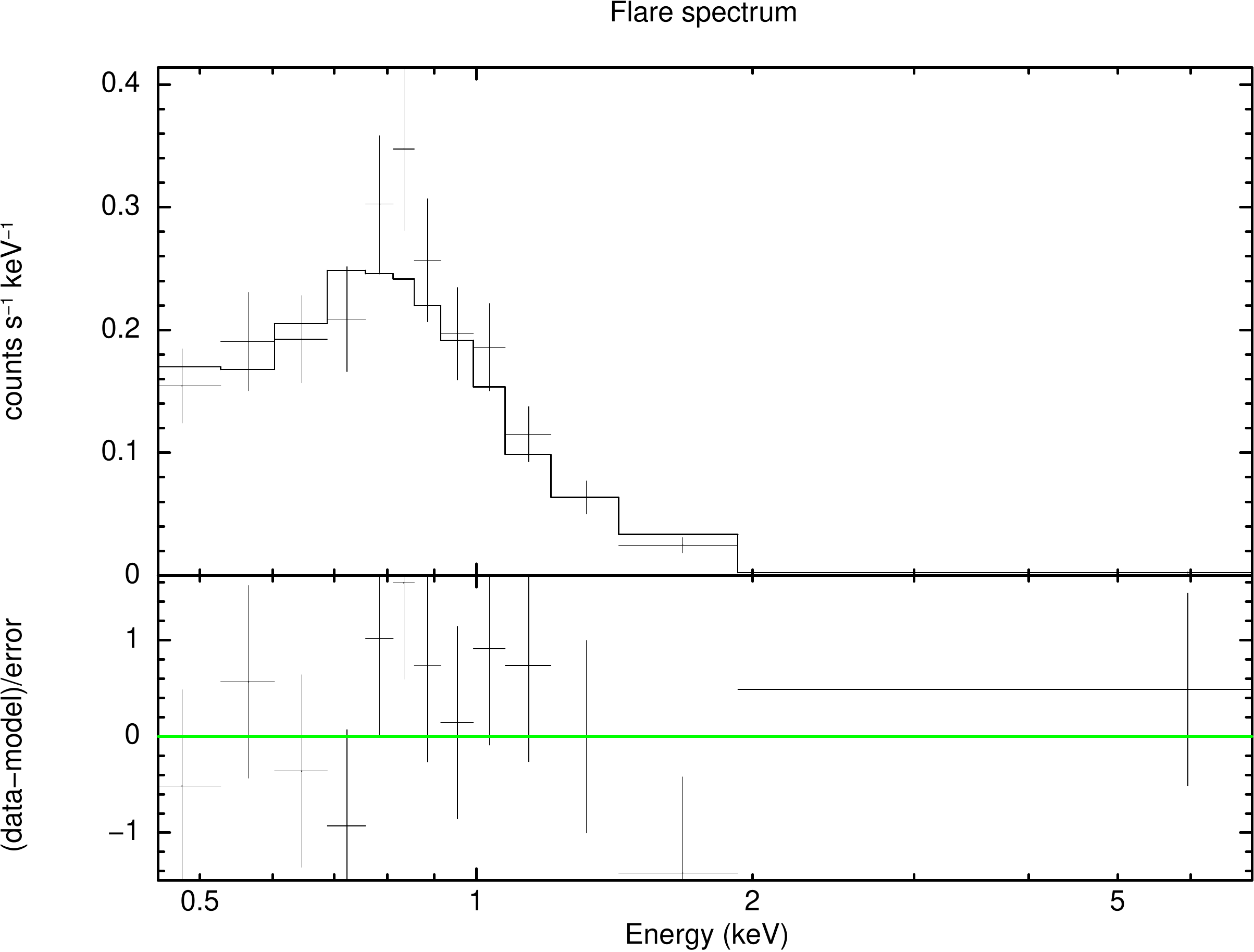}
}
\label{fig:spec}
\caption{\xmm~EPIC spectra of \ktfour\ acquired at different activity phases. 
Left: Spectra obtained in the first (red) and second observation (black) during quiescent intervals.
The bottom panel shows the contributions to $\chi^2$ in units of standard deviations. The model is composed 
of two absorbed APEC thermal components (dotted lines). Right: Same but for the spectrum of \ktfour\  acquired during the flare. }
\end{figure*}

{ In order to improve the count statistics of the spectra, we used the SAS task 
{\sc epicspeccombine} to sum together the spectra of MOS and \pn\ of the same observation and
same time intervals. We refer to the summed spectra as ``EPIC spectra.'' 
The sum was done separately for the quiescent and flaring parts of the light curves.
The quiescent spectrum of the first observation and the whole spectrum of the second
-- quiescent -- observation were fitted simultaneously (Fig. \ref{fig:spec}). 
We also obtained the average spectra of \ktfour\ during each observation ,
accumulating the spectra of the events recorded in each full exposure.

The modeling of the spectra was done with {\sc xspec} version 12.12.1. We used a combination
of APEC
thermal components absorbed by a global equivalent column of H\footnote{In {\sc xspec} syntax: {\sc model tbabs * (cflux*apec + cflux*apec)}}. 
Free parameters were the temperatures, global abundance Z/Z$\odot,$ and fluxes in 0.3--8.0 keV of each component. 
The normalization factors of APEC components were kept fixed to 1 as the { normalization} is
done by the {\it cflux} components. 
The use of {\it cflux} allows a straightforward evaluation of the
unabsorbed flux (with errors) of each APEC component.
We used two APEC components, one for the average and one for the quiescent spectra. 
For the flare spectrum, we used a model composed of the quiescent model with all 
parameters kept fixed and an additional APEC component absorbed by its own equivalent 
H column. This choice allowed us to evaluate any excess of absorption due to material 
ejected during the flare and associated with a coronal mass ejection. For the flare spectrum model, the best-fit parameters were thus n$_H$, kT, and unabsorbed flux.}

The parameters of the best fit for the average, quiescent, and flare phases are
listed in Table \ref{tab:fit}. The errors associated with the best-fit values are 
quoted at the 90\% confidence level. 

\subsection{TESS light curves}
\ktfour\ was observed by TESS  during sectors 14, 20, 21, 40, 41, 47, and 48. For each sector, 
TESS produced an optical light curve with a cadence of two minutes. 
We analyzed these light curves to find different types of variability, 
either impulsive (flares) and modulated or periodic. We adopted the method 
used by \citet{Colombo2022} and based on Gaussian processes and a mixture of different kernels.
In the first step we performed an iterative procedure using the Gaussian process model through the Python package 
{\tt celerite} \citep{Foreman-Mackey2017} with the aim of separating  long-timescale 
activity from the short-timescale events. 
We first modeled the overall long-term changes of the light curve and subtracted this model from the data 
to obtain a curve with median basal flux approximately equal to zero and characterized only by short-timescale events. 
The temporal location of { the peaks of the events and their amplitude} 
were then identified, 
and a fit with a double exponential for the rise and the decay phases was performed. 
The fit determines 
the characteristic times of the rise and the decay when the time of the peak and its amplitude are kept fixed, respectively. 
Lastly, the modeled flares were removed from the light curve, and the residual curve was modeled again with a
quasi-periodic kernel within the {\tt george} Python package \citep{Ambikasaran2014} in order to find any 
modulations due to rotation and/or pulsations. 

\begin{table*}[t]
\caption{\label{tab:fit} Parameters of best-fit models to the average, quiescent, and flare spectra. }
\resizebox{0.95\textwidth}{!}{
\begin{tabular}{lrrrrrrccc}\\ \hline\hline 
 Phase & n$_H$     &  kT$_1$   &   flux$_1$ &  kT$_2$   &    flux$_2$ & Z/Z$\odot$ & $\chi^2$ & D.o.F. & P($\chi^2>\chi^2_0$)\\
  & 10$^{19}$ cm$^{-2}$  & keV &  10$^{-13}$ \fxu & keV &  10$^{-13}$ \fxu &  & & & \% \\ \hline
Average   & 15.3   (0 -- 43.7) & 0.36 (0.32 -- 0.42) & 0.93 (0.79 -- 1.17)  & 0.98 (0.89 -- 1.06) & 0.83 (0.71 -- 0.97) & 0.14 (0.10 -- 0.18) &  117.24 & 94 & 5.3 \\
Quiescent &  3.2   (0 -- 33.9) & 0.35 (0.31 -- 0.40) & 0.82 (0.74 -- 1.07)  & 0.94 (0.86 -- 1.02)   & 0.79 (0.72 -- 0.96) & 0.14 (0.11 -- 0.18) & 100.81 & 84 & 10.2\\
Flare     & $\leq170$ & --  &  --  & 3.5 (1.3 -- 11.8)   & 1.25 (0.94 -- 1.65) & 0.14 (fixed) & 9.35 & 11 & 59.0 \\ \hline
\end{tabular} 
}

\small\smallskip Note: The confidence intervals in parentheses are quoted at the 90\% confidence level. 
Unabsorbed fluxes are given in the 0.3--8.0 keV band. 
\end{table*}

\section{Results \label{results}}
X-ray emission coming from \ktfour\ is detected in both \xmm\ exposures at a
level of 109 $\sigma$ of the local background standard deviation in the combined images
of MOS and \pn\ (Fig. \ref{rgb}).
The light curve shows a flare at t$\sim20$ ks with a duration of about 2~ks.
{ 
We examined the { optical Monitor (OM)} 
light curves obtained in fast mode with the 
UVW1 
filter 
($200-300$ nm). The UV photometry from OM does not show an impulsive variability  during the flare, unlike in X-rays.

Table \ref{tab:fit} reports the best-fit parameters of the spectra during quiescent
and flaring intervals. 
The average spectra from the two full exposures are best fit with 
two APEC thermal components with temperatures of around 0.35 and 1.0 keV, respectively, 
and a light n$_\mathrm H$ absorption on the order of 10$^{20}$ cm$^{-2}$, which is 
consistent with the distance of the star.

The quiescent spectrum is best fit with two components at  
kT$\sim0.35$ keV and $0.94$ keV  and a low but non-negligible gas absorption 
(n$_\mathrm H \sim3.2\times10^{19}$ cm$^{-2}$). 
The quiescent unabsorbed flux is $1.61\times10^{-13}$ \fxu, which gives a luminosity of 
L$_X\sim 1.8\times10^{29}$ \lxu\ at the distance of the star. The ratio $L_X/L_\mathrm{bol} \sim 10^{-5}$ 
suggests that the star is past the saturation phase of X-ray activity seen
in stars younger than Pleiades (100 Myr) and confirms its age to be similar to that of 
Hyades ($\sim700$ Myr). Also,  the quiescent L$_X$ is well within the values observed in Hyades F stars. 
We do not see effects of a magnetic or tidal nature related to SPI  in \ktfour.

{The flare spectrum is best fit with an additional component at $\sim3.5$ keV, although
with a large range of uncertainty (1.27--11.8 keV at the 90\% confidence interval).
During the flare the total flux increases to $2.9\times10^{-13}$ \fxu, resulting in a 
L$_X \sim 3.2\times10^{29}$ \lxu and $L_X/L_\mathrm{bol} \sim 2.2\times10^{-5}$.
Since the softest part of the spectrum below 0.5 keV  has poor count statistics, the n$_H$ absorption for the flaring component is loosely constrained to values $<1.7\times10^{21}$ cm$^{-3}$ and 
consistent with the low value estimated from the quiescent phase. }

 }

\begin{figure}
    \centering
    \resizebox{\columnwidth}{!}{
    \includegraphics{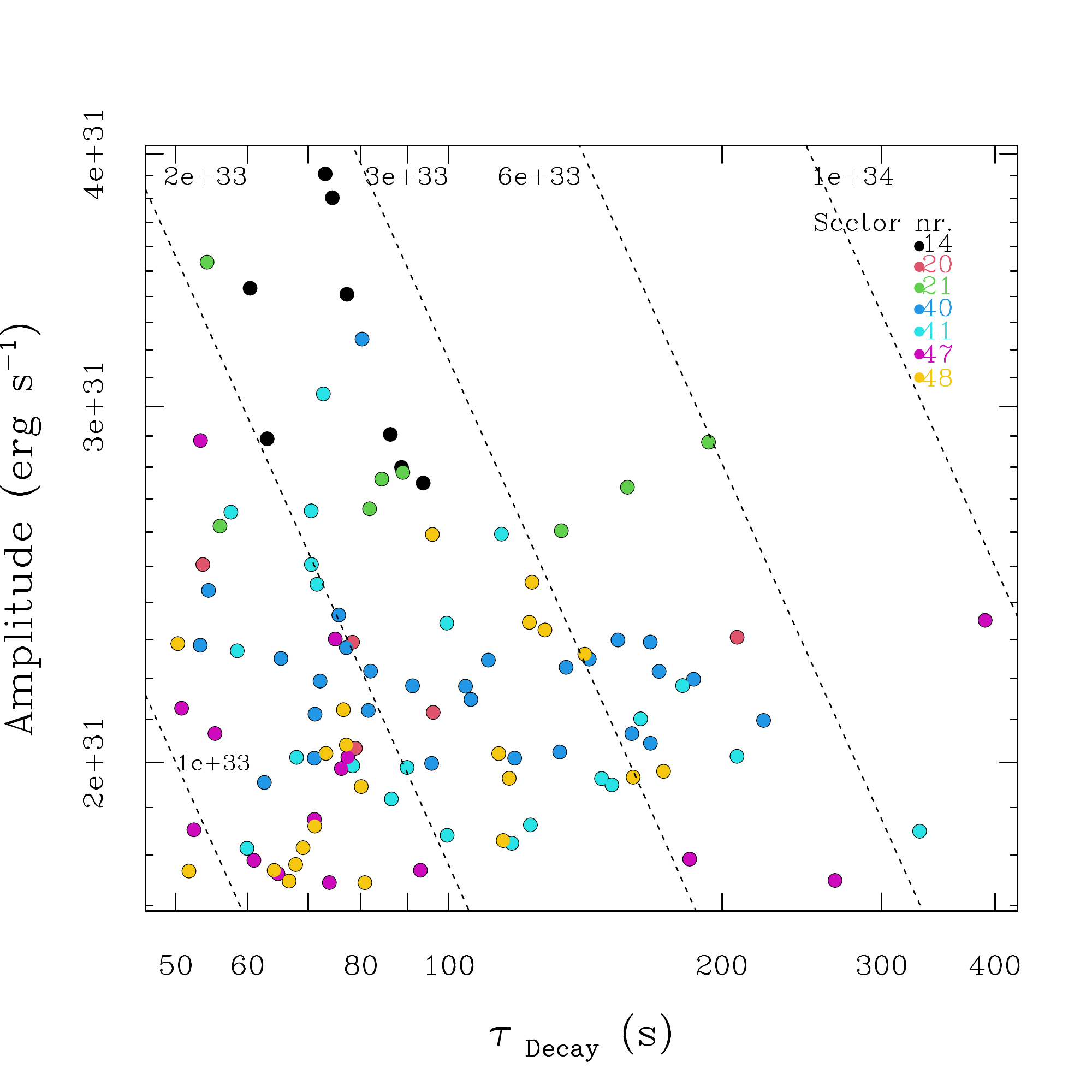}
    }
    \caption{Flare amplitude vs. decay time of the flares detected in TESS. Different colors identify 
    the sectors where the flares were detected. Dashed lines with labels mark the loci of the flares with
    the same energy.}
    \label{fig:tau_amp}
\end{figure}

We detected 108 flares over a total of 187.4 days along seven different TESS sectors.
Figure \ref{fig:tau_amp} shows the peak amplitude luminosity of flares versus their decay time.
The sample shows a number of short-duration flares ($\tau\le100$ s) with high peak 
amplitudes ($L_{flare} \ge>3\times10^{31}$ erg/s) and conversely a number of long-duration flares with
relatively low peak amplitudes. We did not detect long-duration flares ($t\ge200$ s) 
with peak amplitudes larger than$3\times10^{31}$ erg/s.  
We did not find any periodicity of the flares related to either the rotation of the star
or to the orbital period of the planet, and thus no X-ray activity related to SPI.

\begin{figure}
    \centering
    \resizebox{\columnwidth}{!}{
    \includegraphics{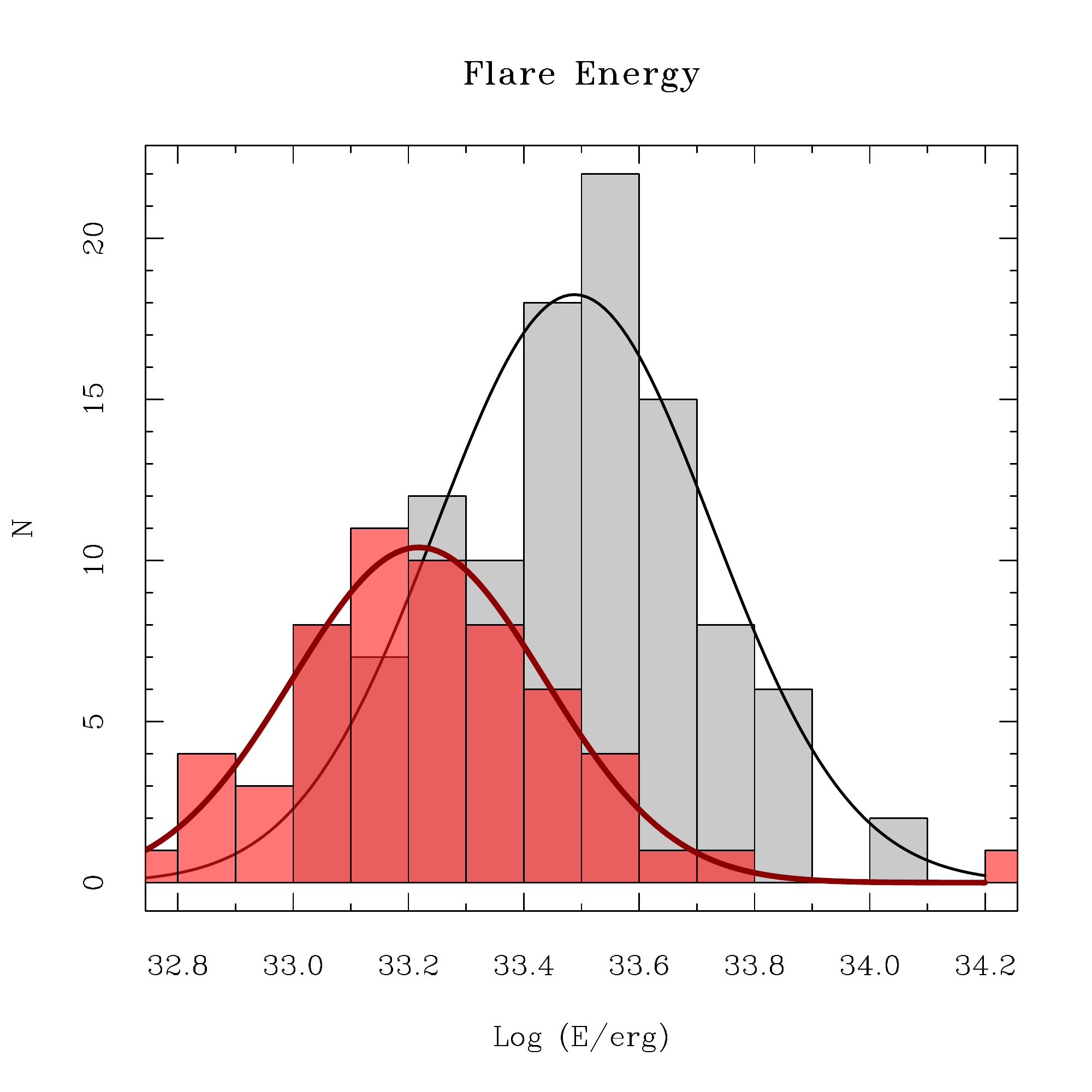}
    }
    \caption{Histogram of energy of the optical flares detected in \ktfour\ (gray histogram) 
    and WASP-18 (red histogram). The histograms are modeled with log-normal distributions 
    (smooth black and red lines) centered at $\log (\mathrm{E/erg}) = 33.48$ for \ktfour\ and $\log 
    (\mathrm{E/erg}) = 33.22$ for WASP-18 and with standard  deviations  of 0.34 dex 0.31 dex, respectively. }
    \label{fig:ex_hist}
\end{figure}

Figure \ref{fig:ex_hist} shows the histogram of the energy of the flares.
{ They were calculated by integrating the flare profiles in TESS light curves
and scaling the bolometric luminosity of the star onto the TESS counts, 
as in \citet[their Eq. 5]{Colombo2022}. }
We modeled the histogram with a log-normal distribution that appears to be centered at around
$\log \mathrm{(E/erg)} \sim 33.48$ and with a standard deviation $\sigma\sim 0.34$ dex. 
These values are not very sensitive to the adopted binning and remain nearly the same when 
using 10 or 15 bins of amplitude values.
The peak of the histogram roughly marks the completeness limit of the sample, so a fraction of flares
with energies below $\log E \sim 33.5$ could remain undetected because of the count statistics 
and noise level of the TESS data. 
This information was used when modeling the cumulative curve of the rates of flares per day. 

\begin{figure}[t]
    \centering
    \resizebox{\columnwidth}{!}{
    \includegraphics{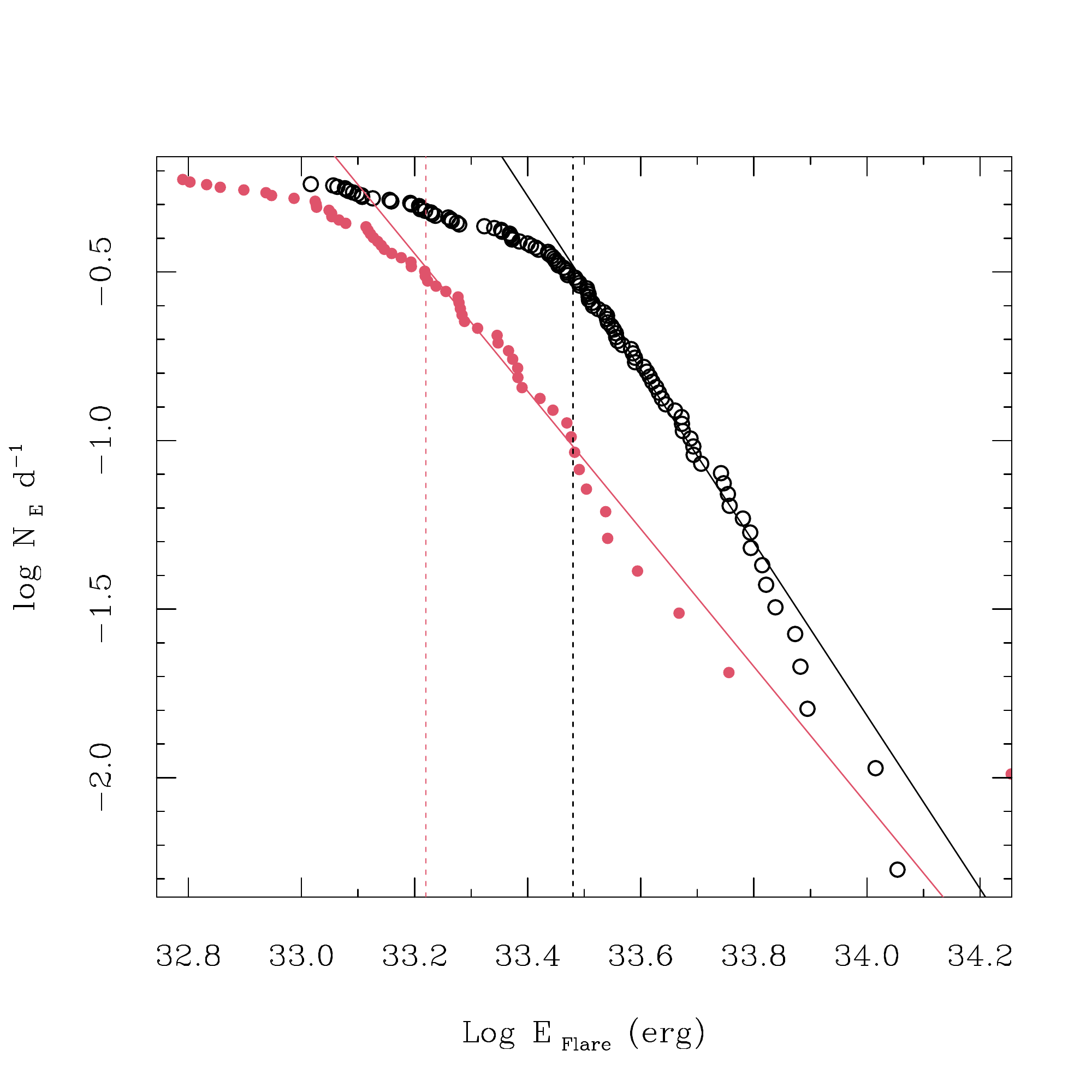}
    }
    \caption{Cumulative distributions of the rate of optical flares detected with TESS 
    as a function of their energies.
    Black points and black lines refer to \ktfour, and red symbols and lines refer to WASP-18.
    The vertical lines mark the peak from the best fit to the histogram of flare energies (cf. Fig. \ref{fig:ex_hist}).
    The fit to the points of the distributions above the knee are marked with solid lines. 
    The slope is about $-2.57\pm0.04$ ($dN/dE \propto E^{-1.57\pm0.04}$) for \ktfour\ and $-2.04\pm0.07$ 
    ($dN/dE \propto E^{-1.04\pm0.07}$) for WASP-18. }
    \label{fig:flare_cumdist}
\end{figure}

We built the cumulative distribution of the rate of flares per day above a given energy detected in the 
TESS light curves (Fig. \ref{fig:flare_cumdist}). The distribution of the rate  of flares shows a "knee" 
or flattening at energies below $\log (E/erg) \le33.4,$ 
which sets the completeness limit of the sample.  We modeled the high energy tail above the knee 
with a linear model in the log-log plane to recover the slope of the distribution and infer the power law index
of the flare energy. To do so, we discarded the last five points for the fit since they are affected by high uncertainties.
From the modeling of $\log N \propto \log E$ of flares above the knee, we infer that $dN/dE\propto E^{-1.57\pm0.04}$.

We applied the same analysis to the TESS light curves of WASP-18, observed in four different sectors, 
in order to detect and measure its flares in the optical band. 
In WASP-18 we detected 58 flares over 97.5 days for an average rate 
of 0.59 flares per day, similar to the rate observed in \ktfour\  (0.57 flares per day). 
The range of the energies of the flares in WASP-18 is, however, lower, in the range 
$32.8 < \log E_\mathrm{opt}/erg< 34.2$ with a median of $\log E_\mathrm{opt}/erg \sim33.22$ (see Fig. \ref{fig:ex_hist}).
The slope of the cumulative distribution of the rate of flares is $dN/dE\propto E^{-1.04\pm0.07}$, 
thus flatter than that measured for \ktfour.

\section{Discussion and conclusions \label{conclusions}}
We investigated the X-ray emission of \ktfour\ with two \xmm\ observations for a total exposure of about 43 ks.
Our aim was to measure the X-ray emission of \ktfour\ and compare it with that of WASP-18.
These two stars both host a massive hJ, { but in WASP-18 the X-ray activity is peculiarly absent.}
\ktfour\ is detected in X-rays and its emission is similar to that of the  
coeval F stars in Hyades. It is characterized by a mildly soft spectrum with a mean thermal component of around 
0.65 keV, a luminosity in 0.3-8 keV of about $\sim1.8\times10^{29}$ \lxu,  and a 
ratio $L_X / L_{bol} \simeq 10^{-5}$.
Altogether, these properties describe the typical level of activity and emission in X-rays of F-type stars 
at about 700 Myr { and are similar to those of F stars in Hyades.}

The analysis of TESS light curves allowed us to determine the frequency of flares as a function of their energy.
With a best fit to the cumulative distribution of flare frequency above the completeness limit of our sample 
(estimated { as} 
$\log E (erg) \sim 33.5$), we estimate that the flares are distributed as 
$dN/dE\propto E^{-1.57\pm0.04}$. 
In the Sun and other stars, a power law relationship is observed for the distribution of flare energy in both optical and X-ray bands with index $dN/dE\propto E^{-2}$
A steeper distribution would indicate a significant contribution of nano-flares that 
could be responsible for the coronal heating. 
In the case of \ktfour,\ the flatter index could mean that a lower fraction of undetected 
small flares and nano-flares are present with respect to the Sun; it would also mean that, in F stars, the relative contribution 
of small flares and nano-flares to the overall variability of their coronae is lower with respect 
to G and K stars.

We detected a flare in X-rays that released an energy of $E_X\sim 1.2\times10^{32}$ erg. 
We can estimate  the energy released in the optical band  $
E_{opt}\sim1.1-10\times10^{33}$ erg by using the calibration of 
\citet{Flaccomio2018}. This empirical relationship was obtained from simultaneous observations 
of very young stars in NGC~2264  with CoROT 
and \textit{Chandra} satellites.  
While the young stars of NGC2264  have X-ray emission levels about 100 times higher 
than that measured in \ktfour, 
the range of optical energy so derived for the X-ray flare of \ktfour\ is in good agreement 
with the energies of the optical flares of \ktfour\ measured from TESS light curves.

We did not find evidence of periodicity of the optical flares synced with the orbital period of 
the planet. We  thus conclude that there are no signs of SPI that
could affect the stellar variability{ at the level detectable with the present TESS data.}
A similar analysis conducted on the TESS light curves of WASP-18 shows a small level of 
optical variability on short timescales that can be associated with flares of energies
lower than those of the flares of \ktfour.

\begin{figure}
\resizebox{\columnwidth}{!}{ 
  \includegraphics{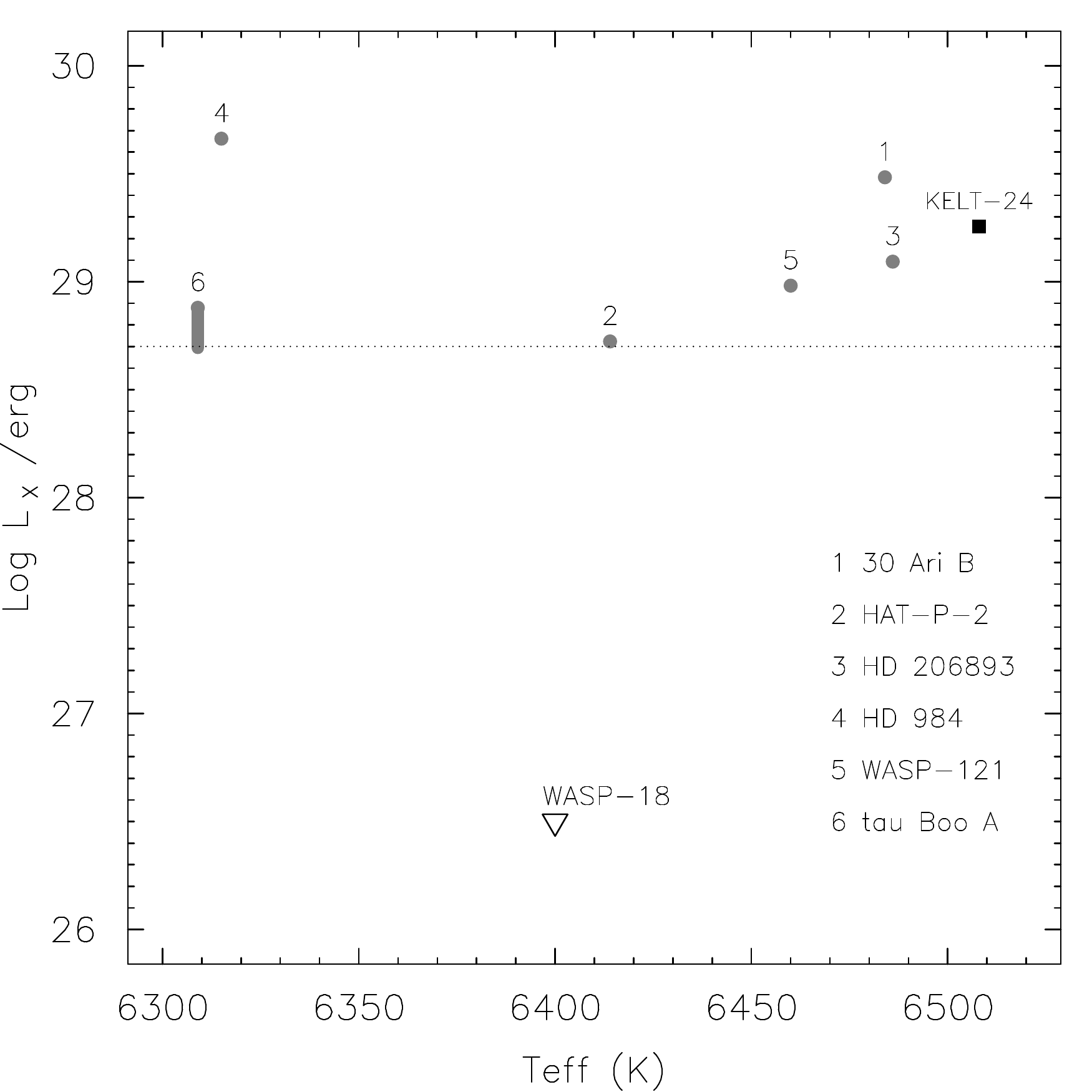} 
  }
\caption{\label{lxfstars} X-ray luminosities in 0.3-8.0 keV vs. T$_\mathrm{eff}$ of a sample of F stars similar to \ktfour\  (star symbol) and WASP-18 (reversed open triangle). 
The names of the stars are reported in the legend.  We added an upper limit of 26.5 for WASP-18 and 
the value obtained with the present data for \ktfour. \xmm\ observed  Tau Bootis A several times;
for this reason, we indicate a range with a vertical segment.  }
\end{figure}

The observation of \ktfour\ in X-rays has served as a comparison test with WASP-18, 
although the latter is a more extreme and perhaps unique system given 
the high mass of its hJ and 
the very short separation between the star and the planet.
The similar X-ray emission of \ktfour\ to that of other F stars in Hyades evidences even more the peculiar lack of X-ray activity in WASP-18.
Beside \ktfour, we extended our search to X-ray emission among other F stars with T$_{eff}$ 
in the range $6300-6500$ K,  with hJs, and within 300 pc in order to be easily detected in X-rays. 
We found six such stars (namely, 30 Ari B, HAT-P-2, HD~206893, HD~984, WASP-121, 
and Tau Bootis A) to compare with \ktfour\ and WASP-18. 
These six F stars have X-ray counterparts in  the ROSAT and \xmm\ catalogs. 
For HD~206893 and HD~984, we found ROSAT count rates that we 
converted to unabsorbed fluxes in the 0.3-8.0 keV band using the
{PIMMS} software. We used a single APEC model of thermal emission from plasma optically thin at 0.54 keV, 
solar coronal abundances, and null gas absorption given the short distances. 
For the rest of the sample we used \xmm\ EPIC fluxes in the $0.2-12$ keV band converted to the $0.3-8.0$ keV
band with PIMMS. Figure \ref{lxfstars} show the luminosities of this sample, to which we added an upper 
limit of 26.5  for the luminosity of WASP-18 and the value of $\log L_\mathrm X = 29.26$ for \ktfour. 
Among this sample of F stars, WASP-18 in the only one undetected in X-rays at a level of about two orders of magnitude lower than that of other F stars. This is at odds with the small optical variability or flickering similar to flares found in the TESS light curves of WASP-18. 
While in the optical band WASP-18 shows some faint activity, in X-rays WASP-18 
is fairly quiet and inactive. 

In \citet{Pillitteri2014} we put forward the idea that the massive very close planet of the WASP-18 system could
tidally interact with its star in a way that hampers the formation of a magnetic activity. 
The lack of magnetism would result in a lack of both chromospheric activity and  X-ray emission.
This hypothesis requires a modeling of the dynamo of the star coupled with the tidal perturbation of
the planet.

Other causes that could make WASP-18 dark in X-rays without SPI 
are that the star is much older than estimated and/or the lack of a 
convective zone inside the star where the magnetic dynamo is formed. 
The age of WASP-18 has been investigated in detail by \cite{Fossati2018}, and their conclusion 
is that  its age appears to be similar  to that of the Hyades cluster: 700-800 Myr. 
At that age, the coronal activity is still present.
The lack of a convective zone would suggest a stellar metallicity much lower than the solar one since the conditions for establishing the convection are never met in the inner layers of the star,
{ similarly to A stars that are dark in X-rays}. 
This does not seem to be the case for WASP-18, which has a metallicity slightly higher than the solar one
([Fe/H]=$+0.11\pm0.08$, \citealp{Torres2012}). Both explanations thus appear not viable. 

In a different scenario, WASP-18 would have entered a prolonged minimum of activity similar 
to the solar Maunder minimum.
\citet{GomesdaSilva2021} studied the chromospheric indicator of activity,  \logrhk, for a sample of solar-type 
stars from F to K observed with the HARPS spectrograph
The distribution of \logrhk for the F stars is characterized 
by a main peak at around -5.0 and a secondary peak at -4.6. 
In this distribution, WASP-18, with \logrhk $\sim -5.4,$ sits at the very low activity tail.  
\citet{GomesdaSilva2021} recognized a group of very inactive stars that could be at
their Maunder minima, including F stars; however, they concluded that they are probably stars 
that have started their evolution off of the main sequence and have thus lost their activity. 
Yet, this seems not to be the case for WASP-18 given its age ($t<1$ Gyr). 

\citet{Baum2022} studied stars with very low activity and reported at least one case where
the activity has decreased below a certain threshold, where the star entered a status similar to the
solar Maunder minimum. WASP-18 could then be another noticeable case in which the stellar activity
has diminished below a critical threshold and thus a long-term minimum of activity has started.
If this is the case, WASP-18 as a peculiar system deserves an accurate monitoring of its activity, either in chromospheric lines sensitive to magnetic activity or in X-rays.

\begin{acknowledgements}
IP,  SC and GM acknowledge financial support from the ASI-INAF agreement n.2018-16-HH.0, 
and from the ARIEL ASI-INAF agreement n.2021-5-HH.0. 
SJW\ was supported by the Chandra X-ray Observatory Center, which is operated by the Smithsonian 
Astrophysical Observatory for and on behalf of the National Aeronautics Space Administration 
under contract NAS8-03060.
Based on observations obtained with XMM-Newton, an ESA science mission with instruments and contributions 
directly funded by ESA Member States and NASA. 
This paper includes data collected by the TESS mission, which are publicly available from the Mikulski 
Archive for Space Telescopes (MAST).
\end{acknowledgements}

%

\begin{thebibliography}{28}
\expandafter\ifx\csname natexlab\endcsname\relax\def\natexlab#1{#1}\fi

\bibitem[{{Ambikasaran} {et~al.}(2014){Ambikasaran}, {Foreman-Mackey},
  {Greengard}, {Hogg}, \& {O'Neil}}]{Ambikasaran2014}
{Ambikasaran}, S., {Foreman-Mackey}, D., {Greengard}, L., {Hogg}, D.~W., \&
  {O'Neil}, M. 2014

\bibitem[{{Baum} {et~al.}(2022){Baum}, {Wright}, {Luhn}, \&
  {Isaacson}}]{Baum2022}
{Baum}, A.~C., {Wright}, J.~T., {Luhn}, J.~K., \& {Isaacson}, H. 2022, \aj,
  163, 183

\bibitem[{{Boller} {et~al.}(2016){Boller}, {Freyberg}, {Tr{\"u}mper}, {Haberl},
  {Voges}, \& {Nandra}}]{Boller2016}
{Boller}, T., {Freyberg}, M.~J., {Tr{\"u}mper}, J., {et~al.} 2016, \aap, 588,
  A103

\bibitem[{{Colombo} {et~al.}(2022{\natexlab{a}}){Colombo}, {Petralia}, \&
  {Micela}}]{Colombo2022}
{Colombo}, S., {Petralia}, A., \& {Micela}, G. 2022{\natexlab{a}}, \aap, 661,
  A148

\bibitem[{{Colombo} {et~al.}(2022{\natexlab{b}}){Colombo}, {Pillitteri},
  {Orlando}, \& {Micela}}]{Colombo2022AN}
{Colombo}, S., {Pillitteri}, I., {Orlando}, S., \& {Micela}, G.
  2022{\natexlab{b}}, Astronomische Nachrichten, 343, e10096

\bibitem[{{Cuntz} {et~al.}(2000){Cuntz}, {Saar}, \& {Musielak}}]{Cuntz2000}
{Cuntz}, M., {Saar}, S.~H., \& {Musielak}, Z.~E. 2000, \apjl, 533, L151

\bibitem[{{Flaccomio} {et~al.}(2018){Flaccomio}, {Micela}, {Sciortino}, {Cody},
  {Guarcello}, {Morales-Calder{\`o}n}, {Rebull}, \& {Stauffer}}]{Flaccomio2018}
{Flaccomio}, E., {Micela}, G., {Sciortino}, S., {et~al.} 2018, \aap, 620, A55

\bibitem[{{Foreman-Mackey} {et~al.}(2017){Foreman-Mackey}, {Agol}, {Angus}, \&
  {Ambikasaran}}]{Foreman-Mackey2017}
{Foreman-Mackey}, D., {Agol}, E., {Angus}, R., \& {Ambikasaran}, S. 2017, ArXiv

\bibitem[{{Fossati} {et~al.}(2018){Fossati}, {Koskinen}, {France}, {Cubillos},
  {Haswell}, {Lanza}, \& {Pillitteri}}]{Fossati2018}
{Fossati}, L., {Koskinen}, T., {France}, K., {et~al.} 2018, \aj, 155, 113

\bibitem[{{Gomes da Silva} {et~al.}(2021){Gomes da Silva}, {Santos},
  {Adibekyan}, {Sousa}, {Campante}, {Figueira}, {Bossini}, {Delgado-Mena},
  {Monteiro}, {de Laverny}, {Recio-Blanco}, \& {Lovis}}]{GomesdaSilva2021}
{Gomes da Silva}, J., {Santos}, N.~C., {Adibekyan}, V., {et~al.} 2021, \aap,
  646, A77

\bibitem[{{Hjorth} {et~al.}(2019){Hjorth}, {Albrecht}, {Talens}, {Grundahl},
  {Justesen}, {Otten}, {Antoci}, {Dorval}, {Foxell}, {Fredslund Andersen},
  {Murgas}, {Palle}, {Stuik}, {Snellen}, \& {Van Eylen}}]{Hjorth2019}
{Hjorth}, M., {Albrecht}, S., {Talens}, G.~J.~J., {et~al.} 2019, \aap, 631, A76

\bibitem[{{Ilic} {et~al.}(2022){Ilic}, {Poppenhaeger}, \&
  {Hosseini}}]{Ilic2022}
{Ilic}, N., {Poppenhaeger}, K., \& {Hosseini}, S.~M. 2022, \mnras, 513, 4380

\bibitem[{{Klein} {et~al.}(2022){Klein}, {Zicher}, {Kavanagh}, {Nielsen},
  {Aigrain}, {Vidotto}, {Barrag{\'a}n}, {Strugarek}, {Nicholson}, {Donati}, \&
  {Bouvier}}]{Klein2022}
{Klein}, B., {Zicher}, N., {Kavanagh}, R.~D., {et~al.} 2022, \mnras, 512, 5067

\bibitem[{{Lanza}(2013)}]{Lanza2013}
{Lanza}, A.~F. 2013, \aap, 557, A31

\bibitem[{{Maggio} {et~al.}(2015){Maggio}, {Pillitteri}, {Scandariato},
  {Lanza}, {Sciortino}, {Borsa}, {Bonomo}, {Claudi}, {Covino}, {Desidera},
  {Gratton}, {Micela}, {Pagano}, {Piotto}, {Sozzetti}, {Cosentino}, \&
  {Maldonado}}]{Maggio2015}
{Maggio}, A., {Pillitteri}, I., {Scandariato}, G., {et~al.} 2015, \apjl, 811,
  L2

\bibitem[{{Matsakos} {et~al.}(2015){Matsakos}, {Uribe}, \&
  {K{\"o}nigl}}]{Matsakos2015}
{Matsakos}, T., {Uribe}, A., \& {K{\"o}nigl}, A. 2015, ArXiv e-prints
  [\eprint[arXiv]{1503.03551}]

\bibitem[{{Miller} {et~al.}(2012){Miller}, {Gallo}, {Wright}, \&
  {Dupree}}]{Miller2012}
{Miller}, B.~P., {Gallo}, E., {Wright}, J.~T., \& {Dupree}, A.~K. 2012, \apj,
  754, 137

\bibitem[{{Pillitteri} {et~al.}(2011){Pillitteri}, {G{\"u}nther}, {Wolk},
  {Kashyap}, \& {Cohen}}]{Pillitteri2011}
{Pillitteri}, I., {G{\"u}nther}, H.~M., {Wolk}, S.~J., {Kashyap}, V.~L., \&
  {Cohen}, O. 2011, \apjl, 741, L18

\bibitem[{{Pillitteri} {et~al.}(2015){Pillitteri}, {Maggio}, {Micela},
  {Sciortino}, {Wolk}, \& {Matsakos}}]{Pillitteri2015}
{Pillitteri}, I., {Maggio}, A., {Micela}, G., {et~al.} 2015, ArXiv e-prints
  [\eprint[arXiv]{1503.05590}]

\bibitem[{{Pillitteri} {et~al.}(2010){Pillitteri}, {Wolk}, {Cohen}, {Kashyap},
  {Knutson}, {Lisse}, \& {Henry}}]{Pillitteri2010}
{Pillitteri}, I., {Wolk}, S.~J., {Cohen}, O., {et~al.} 2010, \apj, 722, 1216

\bibitem[{{Pillitteri} {et~al.}(2014{\natexlab{a}}){Pillitteri}, {Wolk},
  {Lopez-Santiago}, {G{\"u}nther}, {Sciortino}, {Cohen}, {Kashyap}, \&
  {Drake}}]{Pillitteri2014}
{Pillitteri}, I., {Wolk}, S.~J., {Lopez-Santiago}, J., {et~al.}
  2014{\natexlab{a}}, \apj, 785, 145

\bibitem[{{Pillitteri} {et~al.}(2014{\natexlab{b}}){Pillitteri}, {Wolk},
  {Sciortino}, \& {Antoci}}]{Pillitteri2014a}
{Pillitteri}, I., {Wolk}, S.~J., {Sciortino}, S., \& {Antoci}, V.
  2014{\natexlab{b}}, \aap, 567, A128

\bibitem[{{Pillitteri} {et~al.}(2014{\natexlab{c}}){Pillitteri}, {Wolk},
  {Sciortino}, \& {Antoci}}]{Pillitteri2014b}
{Pillitteri}, I., {Wolk}, S.~J., {Sciortino}, S., \& {Antoci}, V.
  2014{\natexlab{c}}, \aap, 567, A128

\bibitem[{{Poppenhaeger} \& {Wolk}(2014)}]{Poppenhaeger2014}
{Poppenhaeger}, K. \& {Wolk}, S.~J. 2014, \aap, 565, L1

\bibitem[{{Rodriguez} {et~al.}(2019){Rodriguez}, {Eastman}, {Zhou}, {Quinn},
  {Beatty}, {Penev}, {Johnson}, {Cargile}, {Latham}, {Bieryla}, {Collins},
  {Dressing}, {Ciardi}, {Relles}, {Murawski}, {Nishiumi}, {Yonehara},
  {Ishimaru}, {Yoshida}, {Gregorio}, {Lund}, {Stevens}, {Stassun}, {Gaudi},
  {Col{\'o}n}, {Pepper}, {Narita}, {Awiphan}, {Chuanraksasat}, {Benni},
  {Zambelli}, {Garrison}, {Wilson}, {Cornachione}, {Wang}, {Labadie-Bartz},
  {Rodr{\'\i}guez}, {Siverd}, {Yao}, {Bayliss}, {Berlind}, {Calkins},
  {Christiansen}, {Cohen}, {Conti}, {Curtis}, {Depoy}, {Esquerdo}, {Evans},
  {Feliz}, {Fulton}, {Holoien}, {James}, {Jayasinghe}, {Jang-Condell},
  {Jensen}, {Johnson}, {Joner}, {Khakpash}, {Kielkopf}, {Kuhn}, {Manner},
  {Marshall}, {McLeod}, {McCrady}, {Oberst}, {Oelkers}, {Penny}, {Reed},
  {Sliski}, {Shappee}, {Stephens}, {Stockdale}, {Tan}, {Trueblood},
  {Trueblood}, {Villanueva}, {Wittenmyer}, \& {Wright}}]{Rodriguez2019}
{Rodriguez}, J.~E., {Eastman}, J.~D., {Zhou}, G., {et~al.} 2019, \aj, 158, 197

\bibitem[{{Shkolnik} {et~al.}(2008){Shkolnik}, {Bohlender}, {Walker}, \&
  {Collier Cameron}}]{Shkolnik2008}
{Shkolnik}, E., {Bohlender}, D.~A., {Walker}, G.~A.~H., \& {Collier Cameron},
  A. 2008, \apj, 676, 628

\bibitem[{{Shkolnik} {et~al.}(2003){Shkolnik}, {Walker}, \&
  {Bohlender}}]{Shkolnik03}
{Shkolnik}, E., {Walker}, G.~A.~H., \& {Bohlender}, D.~A. 2003, \apj, 597, 1092

\bibitem[{{Torres} {et~al.}(2012){Torres}, {Fischer}, {Sozzetti}, {Buchhave},
  {Winn}, {Holman}, \& {Carter}}]{Torres2012}
{Torres}, G., {Fischer}, D.~A., {Sozzetti}, A., {et~al.} 2012, \apj, 757, 161

\end{thebibliography}
%


\end{document}